\documentclass[10pt]{blois07}
\usepackage{wrapfig,psfrag,epsfig,graphicx,graphics,xcolor}
\usepackage[small]{caption2}
\usepackage{amssymb}
\usepackage{amsmath}
\usepackage{amsfonts}
\usepackage{cite,./mcite}

\def\lsim{\raise0.3ex\hbox{$<$\kern-0.75em\raise-1.1ex\hbox{$\sim$}}}
\def\gsim{\raise0.3ex\hbox{$>$\kern-0.75em\raise-1.1ex\hbox{$\sim$}}}
\def\noi{\noindent}

\def\bei{\begin{itemize}}
\def\ei{\end{itemize}}
\def\bea{\begin{eqnarray}}
\def\eea{\end{eqnarray}}
\def\beas{\begin{eqnarray*}}
\def\eeas{\end{eqnarray*}}
\def\beqas{\begin{eqnarray*}}
\def\eqas{\end{eqnarray*}}
\def\beq{\begin{equation}}
\def\eq{\end{equation}}
\def\eeq{\end{equation}}
\def\beqd{\begin{displaymath}}
\def\eeqd{\end{displaymath}}
\def\eqd{\end{displaymath}}

\def\beeq{\begin{eqnarray}} \def\eeeq{\end{eqnarray}}

\def\bef{\begin{frame}}

\def\slashchar#1{\setbox0=\hbox{$#1$}
   \dimen0=\wd0
   \setbox1=\hbox{/} \dimen1=\wd1
   \ifdim\dimen0>\dimen1
      \rlap{\hbox to \dimen0{\hfil/\hfil}}
      #1
   \else
      \rlap{\hbox to \dimen1{\hfil$#1$\hfil}}
      /
   \fi}

\newcommand{\scalingm}{2.2cm}

\newcommand{\scalingml}{2.6cm}
\newcommand{\scaling}{2.98cm}
\newcommand{\scalingb}{3.2cm}
\newcommand{\scalingl}{4cm}
\newcommand{\scalingL}{5cm}
\newcommand{\scalingA}{2.3cm}
\newcommand{\scalingB}{2.8cm}
\newcommand{\scalingC}{3.2cm}
\newcommand{\scalingD}{2.9cm}

\def\ve{\vspace{-.4cm}}

\newcommand{\wids}{0.45\columnwidth}

\newcommand{\lu}{$\ell_1$}
\newcommand{\ld}{$\ell_2$}
\newcommand{\lut}{$\tilde{\ell}_1$}

\newcommand{\rb}{\underline{r}}
\newcommand{\kb}{\underline{k}}

\newcommand{\fin}{\end{document}}

\usepackage[english]{babel}

\usepackage[latin1]{inputenc}

\usepackage{times}
\usepackage[T1]{fontenc}

\begin{document}
\title{Studying QCD factorizations in exclusive $\gamma^* \gamma^* \to \rho^0_L \rho^0_L$}
\author{S.~Wallon$^1$\protect\footnote{ \, talk presented at EDS07},
 B.~Pire$^2$, M.~Segond$^1$, L.~Szymanowski$^{3}$}
\institute{$^1$\,LPT, 
Universit\'e
Paris-Sud, CNRS, Orsay, France,\\ 
$^2$\,CPHT, 
\'Ecole Polytechnique, CNRS, Palaiseau, France, \\ 
$^3$\,IFPA, Universit\'e  de Li\`ege,  
Belgium and SINS,  Warsaw, Poland}
\maketitle

\begin{abstract}
The exclusive process $e^+ \, e^- \to e^+ \, e^-  \, \rho^0_L \, \rho^0_L$ allows  to study 
various dynamics and factorization properties of perturbative QCD. At moderate energy,
we  demonstrate how
collinear QCD factorization emerges, involving generalized
distribution amplitudes (GDA) and transition distribution amplitudes (TDA).
At higher energies,  in the Regge limit of QCD, we
 show that it offers a promising probe of the BFKL resummation effects to be studied at the International Linear Collider (ILC). 

\end{abstract}

\section{Introduction: Exclusive processes at high energy QCD}
\subsection{Motivation}

\begin{wrapfigure}{r}{0.23\columnwidth}
\vspace{-1.2cm}
\centerline{\scalebox{1}
{\epsfig{file=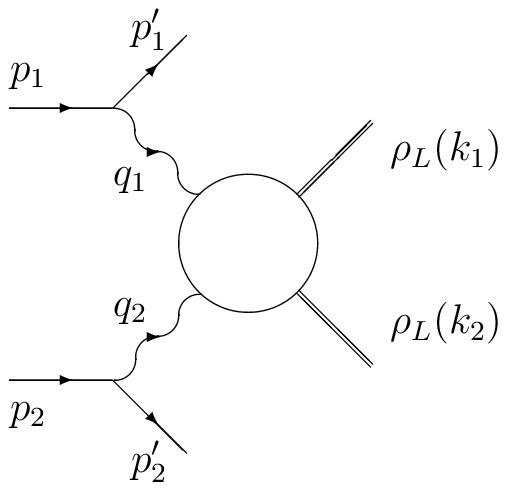,width=3cm}}}\vspace{-.3cm}
\caption{\small  Amplitude for $e^+ \, e^- \to e^+ \, e^-  \, \rho^0_L \, \rho^0_L$.}\label{Figprocessee}
\vspace{-.5cm}
\end{wrapfigure}
Since a decade, there has been much progress in experimental and theoretical  understanding of  hard exclusive processes, including 
Deeply Virtual Compton Scattering (involving Generalized Parton Distributions) and $\gamma \gamma$
scattering
 in fixed target 
   $e^\pm p$ (HERMES, JLab, ...) experiments
and at colliders, such as
   $e^\pm p$  (H1, ZEUS) or 
$e^+ e^-$ (LEP, Belle, BaBar, BEPC).
Meanwhile, the  hard Pomeron \cite{bfkl} concept 
has been developped and tested at
 inclusive (total cross-section), semi-inclusive  (diffraction, forward jets, ...)
and exclusive (meson production) level, for colliders at very large 
energy:
   $e^\pm p$ (HERA), $p \bar{p}$ (Tevatron) and
$e^+ e^-$ (LEP, ILC). Here we  focus on 
\beq
\label{processgg}
\gamma^* \gamma^* \to \rho^0_L \rho^0_L
\eq
with both $\gamma^*$ {\it hard}, through $e^+ e^- \to e^+ e^- \rho^0_L \rho^0_L$
with double tagged outoing leptons (Fig.\ref{Figprocessee}).
It is a beautiful theoretical  laboratory for investigating different {\it dynamics} (collinear, multiregge) and {\it factorization} properties of high energy QCD:
it allows a perturbative study of GPD-like objects at   moderate $s$ and of the hard Pomeron
at  asymptotic $s.$


\subsection{From DIS to GDA and TDA in collinear factorization}

Deep Inelastic Scattering, as an inclusive process, gives access 
to the forward amplitude through the optical theorem. Structure functions 
can be written as 
convolution of  (hard) Coefficient Functions with  (soft) Parton Distributions.  Deeply Virtual Compton Scattering and meson electroproduction on a hadron  $\gamma^*h  \to \gamma \, h, \, h' \, h$,  as exclusive processes, give access to the full amplitude,\nopagebreak which is  a   convolution, for $-t \ll s,$  
of a (hard) CF
with a   (soft) Generalized Parton Distribution\cite{DM,GPD}.
Extensions were made from GPDs.  First\cite{DM,GDA}, the crossed process  $\gamma^* \, \gamma \to h \, h'$ can \nopagebreak be factorized, for  \nopagebreak[4] $s \ll -t,$ as a  convolution of a (hard) CF  with a (soft) Generalized Distribution 
\pagebreak

\psfrag{r1}[cc][cc]{$\quad\rho(k_1)$}
\psfrag{r2}[cc][cc]{$\quad\rho(k_2)$}
\psfrag{p1}[cc][cc]{$\hspace{-1cm}\begin{array}{c} \!\! \ell_1 \hspace{.8cm}\slashchar{p}_1\\  \vspace{-.5cm}\hspace{.2cm}-\tilde{\ell}_1 \end{array}$}
\psfrag{p2}[cc][cc]{$\hspace{-1cm}\begin{array}{c} \!\! \vspace{.2cm}\ell_2 \\  \vspace{.55cm}\hspace{-.4cm}-\tilde{\ell}_2 \hspace{.75cm}\slashchar{p}_2 \end{array}$}
\psfrag{q1}[cc][cc]{$q_1$}
\psfrag{q2}[cc][cc]{$q_2$}
\psfrag{Da}[cc][cc]{DA}
\psfrag{HDA}[cc][cc]{$M_H$}
\psfrag{M}[cc][cc]{$M$}
\begin{wrapfigure}{r}{0.45\columnwidth}
\vspace{-.45cm}
\begin{center}
\hspace{-.4cm}
\scalebox{.63}
{
$\begin{array}{cccc}
\raisebox{-0.44 \totalheight}{\epsfig{file=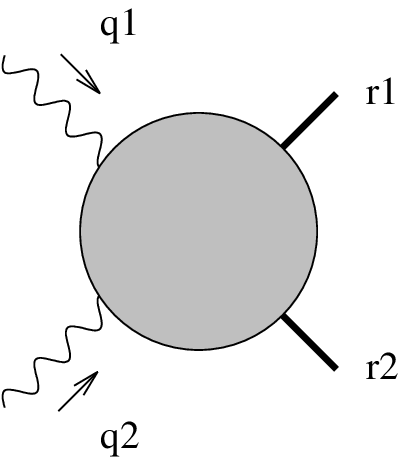,width=\scaling}}&=&
\raisebox{-0.44\totalheight}{\epsfig{file=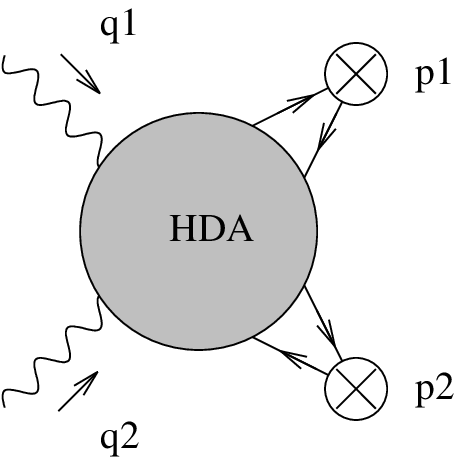,width=\scaling}}
& \begin{array}{c}
\raisebox{0.4 \totalheight}
{\psfrag{r}[cc][cc]{$\quad\rho(k_1)$}
\psfrag{pf}[cc][cc]{$\slashchar{p}_2$}
\epsfig{file=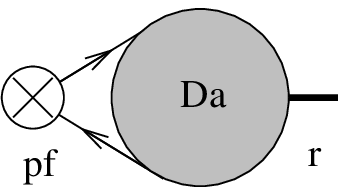,width=\scalingml}}\\
\raisebox{0.1 \totalheight}
{\psfrag{r}[cc][cc]{$\quad\rho(k_2)$}
\psfrag{pf}[cc][cc]{$\slashchar{p}_1$}
\epsfig{file=DA.eps,width=\scalingml}}
\end{array}
\end{array}
$}
\end{center}
\vspace{-.8cm}
\caption{\small 
$\gamma^*(Q_1) \gamma^*(Q_2) \to \rho^0_L (k_{1})\rho^0_L(k_{2})$ with collinear factorization in $q \bar{q} \rho$ vertices.}\vspace{-.7cm}
\label{Figfact}
\end{wrapfigure}
\noi
 Amplitude  describing the correlator between two quark fields and a two hadron state.
Second\cite{TDA}, starting from  meson electroproduction and performing $t \leftrightarrow u$ crossing, and then allowing  the initial and the final hadron  to differ, we write the amplitude for the process $ \gamma^*  \, h \to h" \, h'$ 
as  a  convolution of a  (hard) CF with a (soft) Transition  Distribution Amplitude describing the 
$h \to h'$ transition and with a (soft) Distribution Amplitude (describing   $q \bar{q} h"$ vertex).

\noi We will  rely on
 collinear factorization  for our process (\ref{processgg}) at each  
 $q \bar{q} \rho$ vertex only. At high $Q_i^2$, each of the two quarks
making the $\rho$ mesons are almost collinear, flying in the light cone directions 
$p_1$ and $p_2$ (used as Sudakov vectors), and
 their momentum read
$\ell_i\sim z_i\, k_i$ and
$\tilde{\ell}_i\sim \bar{z}_i\, k_i  \,.$
The amplitude $M$ is  factorized   as a convolution of a hard part $M_H$ with two $\rho^0_L$ DAs (see Fig.\ref{Figfact}),  defined 
as matrix elements of non local quarks fields correlator   on the light 
cone\footnote{We limit ourselves to longitudinally polarized mesons to avoid potential end-point singularities.} \footnote{$\phi(z)=6 z (1-z)\,(1+ \sum_{n=1}^{\infty} a_{2  n} C^{3/2}_{2\,n} (2 z-1))\,.$ } 
\vspace{-.2cm}
$$\vspace{-.7cm}\hspace{-2cm}
\langle \rho^0_L(k)|\bar q(x)\gamma^\mu q(0)|0\rangle =
\frac{f_\rho}{\sqrt{2}}\; k^\mu \;\int\limits_0^1\,dz\,e^{iz(kx)}\phi(z)\;, \;\;\;\;\;\mbox{for}\;\;q=u,d\;.
$$

\section{Computation at fixed $W^2$}

\subsection{Direct calculation}
\begin{wrapfigure}{r}{0.45\columnwidth}
\psfrag{q1}[cc][cc]{$q_1$}
\psfrag{q2}[cc][cc]{$q_2$}
\psfrag{l1}[cc][cc]{\lu}
\psfrag{l1t}[cc][cc]{-\lut}
\psfrag{l2}[cc][cc]{\ld}
\psfrag{l2t}[cc][cc]{$\!-\tilde{\ell}_2$}
\psfrag{n}[cc][cc]{}
\vspace{-1.3cm}
\centerline{\raisebox{1.3 \totalheight}{\scalebox{.4}{$
\begin{array}{cccccccc}
\raisebox{-0.44 \totalheight}{\epsfig{file=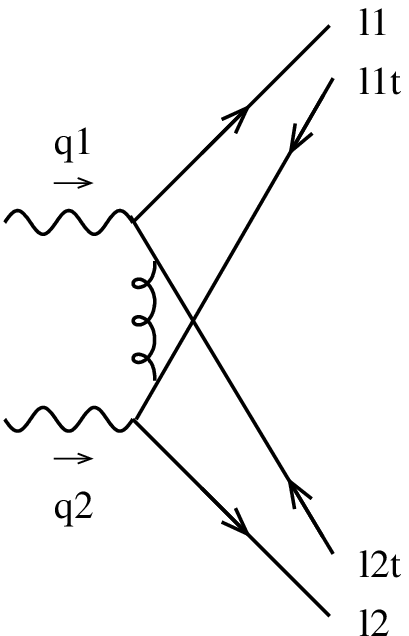,width=\scalingb}}
&+&\raisebox{-0.44 \totalheight}{\epsfig{file=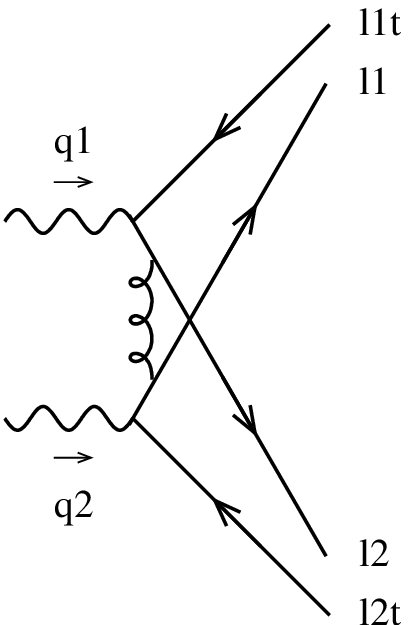,width=\scalingb}} &+&\raisebox{-0.44 \totalheight}{\epsfig{file=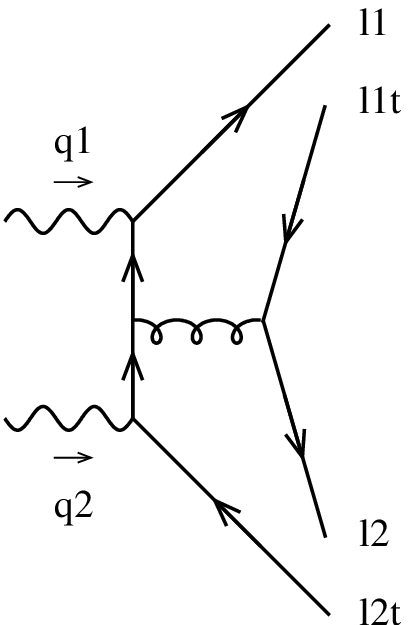,width=\scalingb}}&+&\raisebox{-0.44 \totalheight}{\epsfig{file=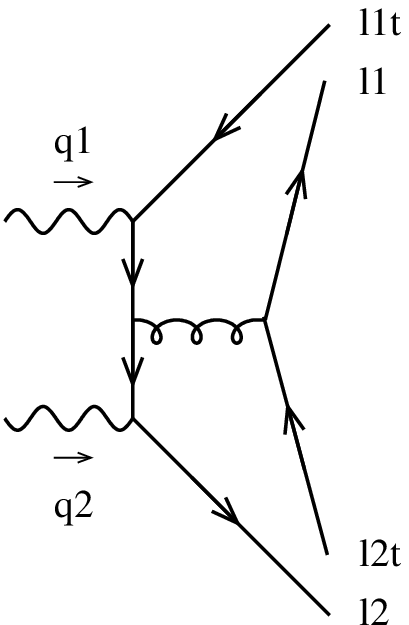,width=\scalingb}}
\end{array}
$}}}
\vspace{-1.8cm}
\caption{Diagrams contributing to $M_H$
for $\gamma^*_L$. \label{longDiagrams}}\vspace{-1.cm}
\end{wrapfigure}



We compute\cite{gdatda} the amplitude $M$ following the Brodsky, Lepage approach\cite{BLphysrev24}. At  Born order (quark exchange) and in the
 forward case for simplicity, 
the amplitude $M$
reads\footnote{$g^{\mu\,\nu}_T=g^{\mu \nu} - \frac{p_1^\mu p_2^\nu + p_1^\nu p_2^\mu}{p_1 . p_2}\,.$}, 
\beqd
M \!=  \!T^{\mu\, \nu}\!\epsilon_\mu(q_1)\epsilon_\nu(q_2)\,, \  T^{\mu\, \nu}  \!= \!\frac{1}{2}g^{\mu\,\nu}_T\;  T^{\alpha\, \beta}g_{T\,\alpha\,\beta}+
\left(\!p_1^\mu +\frac{Q_1^2}{s}p_2^\mu\!\right)\!\left(\!p_2^\nu
+ \frac{Q_2^2}{s}p_1^\nu\!\right)\frac{4}{s^2}  T^{\alpha\, \beta}p_{2\,\alpha}\, p_{1\,\beta}\,.
\eqd
In the case of longitudinally  polarized photons, their polarization vectors read, with $s \equiv 2 \, p_1 \cdot p_2,$
\beq
\label{polL}
\epsilon_\parallel(q_1)=\frac{1}{Q_1}q_1 + \frac{2 Q_1}{s}p_2 \quad {\rm and} \quad \epsilon_\parallel(q_2)=\frac{1}{Q_2}q_2 + \frac{2 Q_2}{s}p_1\,.
\eeq
Due to QED gauge invariance, the first terms in RHS of (\ref{polL}) do  not contribute.
In the forward case discussed here,
 the number of diagrams then reduces to 4, as illustrated in Fig.\ref{longDiagrams}. They result into
\beqd
T^{\alpha\, \beta}p_{2\,\alpha} \,p_{1\,\beta}= -\frac{s^2 f_\rho^2 C_F e^2 g^2(Q_u^2 +Q_d^2)}{8N_{c}Q_1^2 Q_2^2}
\int\limits_0^1\,dz_1\,dz_2\,\phi(z_1)\,\phi(z_2)
\eqd
\nopagebreak[4]
\begin{equation}
\times \left\{\frac{(1-\frac{Q^2_1}{s})(1-\frac{Q^2_2}{s})}{(z_1+\bar z_1 \frac{Q^2_2}{s})(z_2+\bar z_2\frac{Q^2_1}{s})}
+ \frac{(1-\frac{Q^2_1}{s})(1-\frac{Q^2_2}{s})}{(\bar z_1+ z_1 \frac{Q^2_2}{s})(\bar z_2+ z_2\frac{Q^2_1}{s})}
+ \frac{1}{z_2 \bar z_1} + \frac{1}{z_1 \bar z_2} \right\}.\ \ \ \ \
\label{resML}
\end{equation}
\noi
For transversally polarized photons, 
 no simplification occurs and the 12 diagrams give
 \begin{eqnarray}
\vspace{-1cm}
T^{\alpha\, \beta}g_{T\,\alpha \,\beta}&=& -\frac{e^2(Q_u^2 +Q_d^2)\,g^2\,C_F\,f_\rho^2}{4\,N_c\,s}
\int\limits_0^1\,dz_1\,dz_2\,\phi(z_1)\,\phi(z_2)\left\{2\left(1-\frac{Q^2_2}{s}\right)\left(1-\frac{Q^2_1}{s}\right) \right.\nonumber
 \\
&&\hspace{-2cm}\times \left. \left[
\frac{1}{(z_2+\bar z_2\frac{Q_1^2}{s})^2(z_1+\bar z_1\frac{Q_2^2}{s} )^2} +
\frac{1}{(\bar z_2+ z_2\frac{Q_1^2}{s})^2(\bar z_1+ z_1\frac{Q_2^2}{s} )^2}
 \right] \; + \left(\frac{1}{\bar z_2\,z_1}- \frac{1}{\bar z_1\,z_2}  \right) \right.
\nonumber \\
 && \hspace{-2cm}\left.
\times \,
\left[ \frac{1}{1-\frac{Q^2_2}{s}}\left( \frac{1}{\bar z_2+ z_2\frac{Q_1^2}{s}}
                                      -\frac{1}{ z_2+ \bar z_2\frac{Q_1^2}{s}}   \right)
 - \frac{1}{1-\frac{Q^2_1}{s}}\left(\frac{1}{\bar z_1+ z_1\frac{Q_2^2}{s}} - \frac{1}{ z_1+ \bar z_1\frac{Q_2^2}{s}}   \right) \right]
\right\} \,.
\label{resMT}
\end{eqnarray}
The $ z_i$ integrations have no end-point singularity ($Q_i^2$ are non-zero and  DAs vanishes at $z_i=0$).

\subsection{Interpretation in terms of QCD Factorization}
\subsubsection{GDA for transverse photon in the limit $\Lambda_{QCD}^2 \ll W^2 \ll Max(Q_1^2,Q_2^2)$}

When $W^2$ is smaller than the highest photon virtuality,
the  direct calculation (\ref{resMT}) simplifies in\footnote{In
this example
$\frac{W^2}{Q_1^2}=
\frac{s}{Q_1^2}\left(1-\frac{Q_1^2}{s}
\right)\left(1-\frac{Q_2^2}{s}  \right) \approx 1-\frac{Q_1^2}{s}\ll
1;$ we denote $C=\frac{e^2(Q_u^2 +Q_d^2)\,g^2\,C_F\,f_\rho^2}{4\,N_c}$}
\beq
\label{simpMT}
T^{\alpha\, \beta}g_{T\,\alpha \,\beta} \approx \frac{C}{W^2}
\int\limits_0^1\,dz_1\,dz_2\,\left(\frac{1}{\bar z_1+ z_1\frac{Q_2^2}{s}}
 - \frac{1}{ z_1+ \bar z_1\frac{Q_2^2}{s}}   \right)\,
\left(\frac{1}{\bar z_2\,z_1}- \frac{1}{\bar z_1\,z_2}  \right)\, \phi(z_1)\,\phi(z_2)\nonumber
\eq

\psfrag{r1}[cc][cc]{$\quad\rho(k_1)$}
\psfrag{r2}[cc][cc]{$\quad\rho(k_2)$}
\psfrag{p1}[cc][cc]{$\slashchar{p}_1$}
\psfrag{p2}[cc][cc]{$\slashchar{p}_2$}
\psfrag{p}[cc][cc]{$\qquad\slashchar{P}\qquad \slashchar{n}$}
\psfrag{n}[cc][cc]{}
\psfrag{q1}[cc][cc]{$q_1$}
\psfrag{q2}[cc][cc]{$q_2$}
\psfrag{GDA}[cc][cc]{${GDA}_H$}
\psfrag{Da}[cc][cc]{DA}
\psfrag{HDA}[cc][cc]{$M_H$}
\psfrag{M}[cc][cc]{$M$}
\psfrag{Th}[cc][cc]{$T_H$}
\begin{wrapfigure}{r}{0.64\columnwidth}
\vspace{-1.2cm}
\begin{center}
\hspace{-.2cm}
\scalebox{.57}
{$\begin{array}{cccccc}
\!\!\raisebox{-0.44\totalheight}{\epsfig{file=HDA.eps,width=\scalingD}}& \begin{array}{c}
\raisebox{0.51 \totalheight}
{\psfrag{r}[cc][cc]{$\quad\rho(k_1)$}
\psfrag{pf}[cc][cc]{$\slashchar{p}_2$}
\epsfig{file=DA.eps,width=\scalingA}}\\
\raisebox{0. \totalheight}
{\psfrag{r}[cc][cc]{$\quad\rho(k_2)$}
\psfrag{pf}[cc][cc]{$\slashchar{p}_1$}
\epsfig{file=DA.eps,width=\scalingA}}
\end{array}&=& \!
\raisebox{-0.44 \totalheight}
{\psfrag{r}[cc][cc]{$\quad\rho(k_1)$}
\psfrag{pf}[cc][cc]{$\slashchar{p}_2$}
\epsfig{file=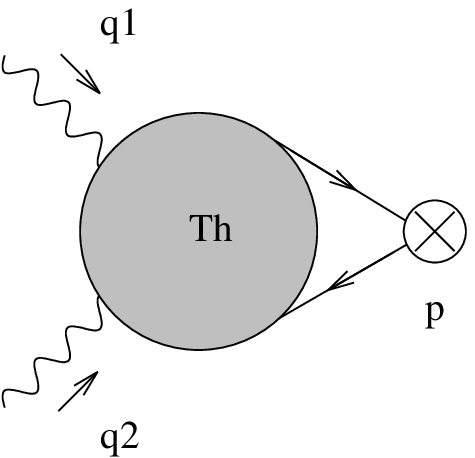,width=\scalingB}}&
\raisebox{-0.43\totalheight}{\epsfig{file=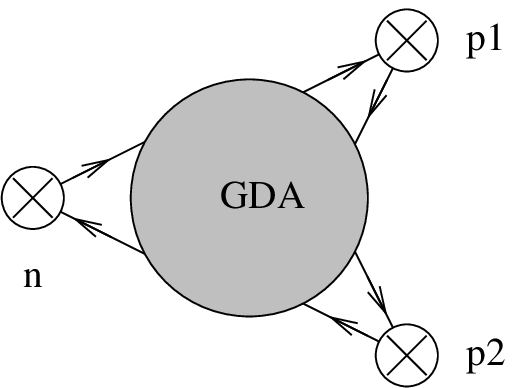,width=\scalingC}}& \begin{array}{c}
\raisebox{0.51 \totalheight}
{\psfrag{r}[cc][cc]{$\quad\rho(k_1)$}
\psfrag{pf}[cc][cc]{$\slashchar{p}_2$}
\epsfig{file=DA.eps,width=\scalingA}}\\
\raisebox{-0.0 \totalheight}
{\psfrag{r}[cc][cc]{$\quad\rho(k_2)$}
\psfrag{pf}[cc][cc]{$\slashchar{p}_1$}
\epsfig{file=DA.eps,width=\scalingA}}
\end{array}
\end{array}
$}
\end{center}
\vspace{-.2cm}
\caption{Factorisation of the  amplitude in terms of a GDA.}
\label{FigconvGDA}
\ve
\end{wrapfigure}
\noi  showing that the hard amplitude $M_H$ can be factorized as a convolution between a hard coefficient function $T_H$ and a ${GDA}_H,$ itself {\it perturbatively} computable (Fig.\ref{FigconvGDA}), extending the results of \cite{DFKV}.
This is proven at Born order by computing perturbatively the GDA  from its definition 
\beqd
 \begin{array}{l}
\langle \rho^0_L(k_1)\,\rho^0_L(k_2)| \bar q(-\alpha \;n/2)\,\slashchar{n}\,
\exp \left[i g \int\limits_{-\frac{\alpha}{2}}^{\frac{\alpha}{2}}\,dy \, n_\nu\, A^\nu (y) \right]
q(\alpha \;n/2)|0\rangle \\
=
 \int\limits_0^1\,dz\,e^{-i(2z-1)\alpha (nP)/2} \Phi^{\rho^0_L\rho^0_L}(z,\zeta,W^2) \quad ({\rm for} \,  Q_1 > Q_2, \, P \sim p_1 \ {\rm and} \ n \sim p_2 )
\end{array}
\eqd

\begin{figure}[h]
\vspace{-.5cm}
\centerline{\scalebox{1.}{\begin{tabular}{ccc}
\scalebox{.52}
{$\begin{array}{cccc}
\raisebox{-0.44 \totalheight}
{\psfrag{r}[cc][cc]{$\quad\rho(k_1)$}
\psfrag{pf}[cc][cc]{$\slashchar{p}_2$}
\epsfig{file=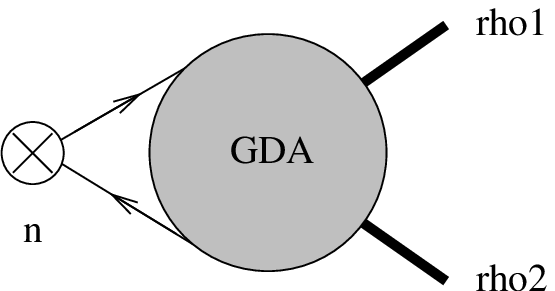,width=\scalingB}}&=&
\raisebox{-0.43\totalheight}{
\psfrag{GDA}[cc][cc]{$GDA_H$}
\epsfig{file=GDA.eps,width=\scalingC}}& \begin{array}{c}
\raisebox{0.51 \totalheight}
{\psfrag{r}[cc][cc]{$\quad\rho(k_1)$}
\psfrag{pf}[cc][cc]{$\slashchar{p}_2$}
\epsfig{file=DA.eps,width=\scalingA}}\\
\raisebox{-0.0 \totalheight}
{\psfrag{r}[cc][cc]{$\quad\rho(k_2)$}
\psfrag{pf}[cc][cc]{$\slashchar{p}_1$}
\epsfig{file=DA.eps,width=\scalingA}}
\end{array}
\end{array}$
}
&\hspace{-.2cm}
 with
&\hspace{-.2cm}
\scalebox{.52}{
$\begin{array}{ccccccc}
\psfrag{GDA}[cc][cc]{$GDA_H$}
\psfrag{p1}[cc][cc]{$\slashchar{p}_1$}
\psfrag{p2}[cc][cc]{$\slashchar{p}_2$}
\psfrag{n}[cc][cc]{$\slashchar{n}$}
\raisebox{-0.43\totalheight}{\epsfig{file=GDA.eps,width=\scalingC}} & =&
\psfrag{p1}[cc][cc]{$\slashchar{p}_1$}
\psfrag{p2}[cc][cc]{$\slashchar{p}_2$}
\psfrag{n}[cc][cc]{$\slashchar{n}$}
\raisebox{-0.44 \totalheight}{\epsfig{file=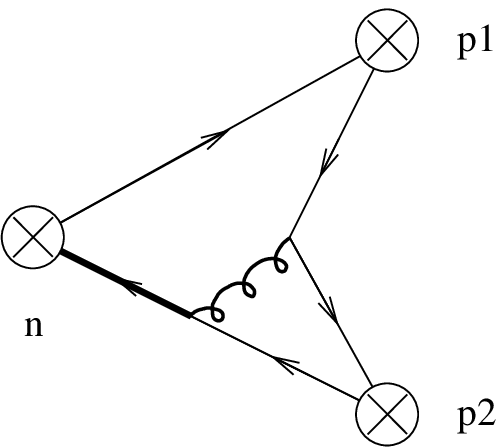,width=\scaling}}
&+&
\psfrag{p1}[cc][cc]{$\slashchar{p}_1$}
\psfrag{p2}[cc][cc]{$\slashchar{p}_2$}
\psfrag{n}[cc][cc]{$\slashchar{n}$}
\raisebox{-0.44\totalheight}{\epsfig{file=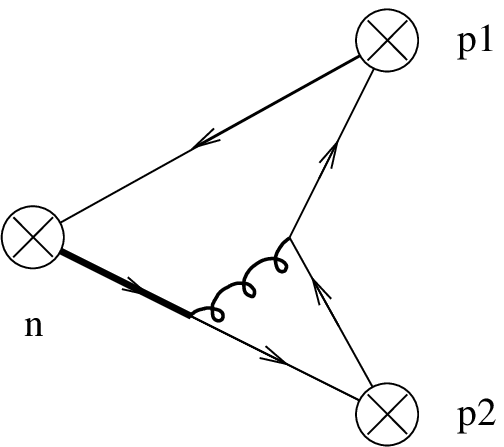,width=\scaling}}&+&
\psfrag{p1}[cc][cc]{$$}
\psfrag{p2}[cc][cc]{$$}
\raisebox{-0.44\totalheight}{\epsfig{file=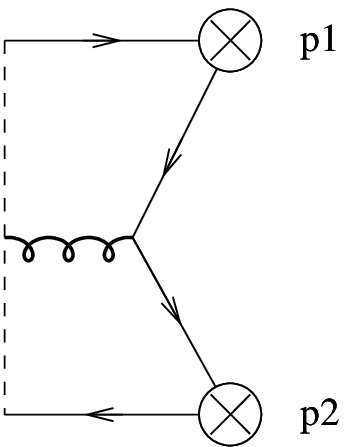,width=2cm}}
\end{array}$
}
\end{tabular}}}\vspace{-.2cm}
\caption{Perturbative GDA factorization.}
\label{FigfactGDAH}
\ve
\end{figure}

\psfrag{s}[cc][cc]{$W^2$}
\psfrag{g}[cc][cc]{$Q_2^2$}
\psfrag{gs}[cc][cc]{$Q_1^2$}
\psfrag{z}[cc][cc]{$z$}
\psfrag{zm}[cc][cc]{$1-z$}
\psfrag{hu}[cc][cc]{$\hspace{-1cm}\zeta \hspace{.7cm} \rho^0_L$}
\psfrag{hd}[cc][cc]{$\hspace{-1.3cm}1-\zeta  \hspace{.5cm}\rho^0_L$}
\psfrag{GDA}[cc][cc]{GDA}
\begin{wrapfigure}{r}{0.3\columnwidth}
\vspace{-1cm}
\centerline{\scalebox{.51}{\raisebox{-0.44\totalheight}{\epsfig{file=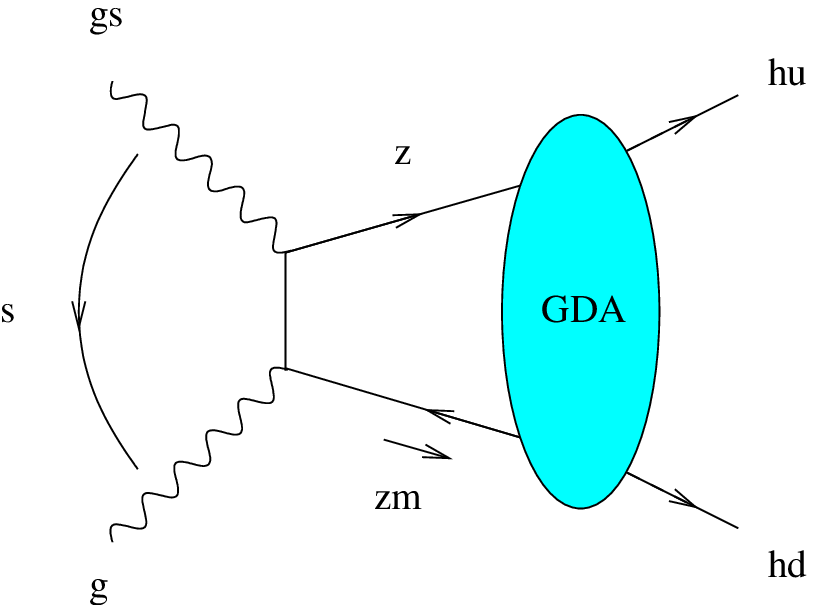,width=\scalingL}}}}
\caption{GDA Kinematics.}
\label{FigcalcGDA}
\vspace{-.8cm}
\end{wrapfigure}
\noi with the kinematics fixed according to Fig.\ref{FigcalcGDA}.  $W^2$ being hard, the GDA can be factorized into Hard part  $\otimes$ DA  DA (see 
Fig.\ref{FigfactGDAH}), as 
\[
\Phi^{\rho_L\rho_L}(z,\zeta\approx 1,W^2)\!
=\! -\frac{f_\rho^2\,g^2\,C_F}{2\,N_c\,W^2}\!\int\limits_0^1\!dz_2\,\phi(z)\,\phi(z_2)
\!\left[\frac{1}{z\bar z_2} \!-\!\frac{1}{\bar z z_2} \right]\,.
\]
\psfrag{r1}[cc][cc]{$\quad\rho(k_1)$}
\psfrag{r2}[cc][cc]{$\quad\rho(k_2)$}
\psfrag{p1}[cc][cc]{$\slashchar{p}_1$}
\psfrag{p2}[cc][cc]{$\slashchar{p}_2$}
\psfrag{p}[cc][cc]{$\qquad\slashchar{P}\qquad \slashchar{n}$}
\psfrag{n}[cc][cc]{}
\psfrag{q1}[cc][cc]{$q_1$}
\psfrag{q2}[cc][cc]{$q_2$}
\psfrag{GDA}[cc][cc]{$GDA$}
\psfrag{Da}[cc][cc]{DA}
\psfrag{HDA}[cc][cc]{$M_H$}
\psfrag{M}[cc][cc]{$M$}
\psfrag{Th}[cc][cc]{$T_H$}
\psfrag{rho1}[cc][cc]{$\quad\rho(k_1)$}
\psfrag{rho2}[cc][cc]{$\quad\rho(k_2)$}

\psfrag{p}[cc][cc]{$\slashchar{P}$}
\psfrag{q1}[cc][cc]{$q_1$}
\psfrag{q2}[cc][cc]{$q_2$}
\psfrag{Th}[cc][cc]{$T_H$}
\psfrag{n}[cc][cc]{$\slashchar{n}$}
\begin{wrapfigure}{r}{0.38\columnwidth}
\vspace{-.1cm}
\centerline{\scalebox{.48}
{
$\begin{array}{ccccc}
\raisebox{-0.46 \totalheight}{\epsfig{file=THT.eps,width=\scaling}}
&=&
\raisebox{-0.46 \totalheight}{\epsfig{file=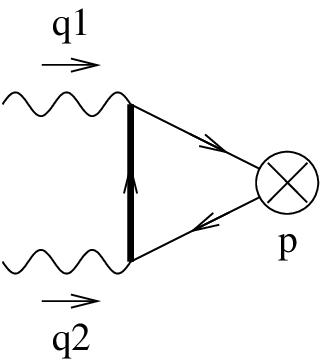,width=\scaling}}
&+&
\raisebox{-0.46\totalheight}{\epsfig{file=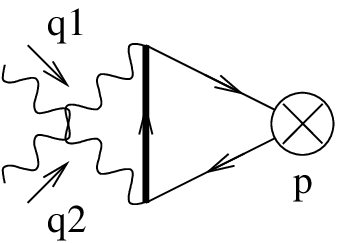,width=\scaling}}
\end{array}
$}}
\caption{Hard part $T_H$ at lowest order.} 
\label{FigTHGDA}\vspace{-.5cm}
\end{wrapfigure}
\noi In  forward kinematics, the QCD Wilson line (last term in  Fig.\ref{FigfactGDAH}) vanishes. The Born order hard part 
is (see Fig.\ref{FigTHGDA})
\[
T_H(z) = -4\,e^2\,N_c\,Q_q^2\,\left( \frac{1}{\bar z +z\,\frac{Q_2^2}{s}} -
   \frac{1}{z + \bar z\,\frac{Q_2^2}{s}} \right)\,.
\]



\subsubsection{TDA for longitudinal photon in the limit $Q_1^2 \gg Q_2^2$ (or $Q_1^2 \ll Q_2^2$)}

\begin{wrapfigure}{r}{1.\columnwidth}
\psfrag{q1}[cc][cc]{$q_1$}
\psfrag{q2}[cc][cc]{$q_2$}
\psfrag{p1}[cc][cc]{$\slashchar{p}_1$}
\psfrag{p2}[cc][cc]{$\slashchar{p}_2$}
\psfrag{n}[cc][cc]{$\slashchar{p}_1$}
\psfrag{p}[cc][cc]{$\slashchar{p}_2$}
\psfrag{Tda}[cc][cc]{$TDA_H$}
\psfrag{Th}[cc][cc]{$T_H$}
\psfrag{Da}[cc][cc]{DA}
\vspace{-.4cm}
\hspace{1cm}\begin{tabular}{cc}
\scalebox{.6}
{
$\begin{array}{c}
\begin{array}{cc}
\raisebox{-0.44 \totalheight}{\epsfig{file=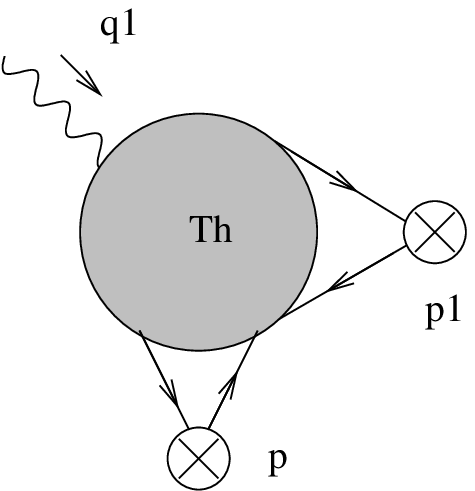,width=\scaling}}&
\raisebox{-0.25\totalheight}
{\psfrag{r}[cc][cc]{$\qquad \rho(k_1)$}
\psfrag{pf}[cc][cc]{\raisebox{-1.05 \totalheight}{$\slashchar{p}_2$}}\epsfig{file=DA.eps,width=\scalingm}}
\end{array}\\
\\
\begin{array}{cc}
\raisebox{-0.44 \totalheight}{\epsfig{file=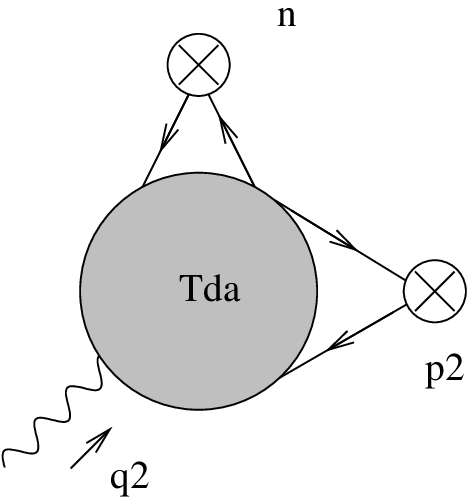,width=\scaling}}&
\raisebox{-0.54\totalheight}
{\psfrag{r}[cc][cc]{$\qquad \rho(k_2)$}
\psfrag{pf}[cc][cc]{\raisebox{-1.05 \totalheight}{$\slashchar{p}_1$}}\epsfig{file=DA.eps,width=\scalingm}}
\end{array}
\end{array}
$}
& \hspace{1cm}
\begin{tabular}{c}
\psfrag{r1}[cc][cc]{$\quad\rho(k_1)$}
\psfrag{r2}[cc][cc]{$\quad\rho(k_2)$}
\psfrag{p1}[cc][cc]{$\slashchar{p}_1$}
\psfrag{p2}[cc][cc]{$\slashchar{p}_2$}
\psfrag{p}[cc][cc]{$\qquad\slashchar{P}\qquad \slashchar{n}$}
\psfrag{n}[cc][cc]{$\slashchar{p}_1$}
\psfrag{q1}[cc][cc]{$q_1$}
\psfrag{q2}[cc][cc]{$q_2$}
\psfrag{Tda}[cc][cc]{$TDA$}
\psfrag{Da}[cc][cc]{DA}
\psfrag{HDA}[cc][cc]{$M_H$}
\psfrag{M}[cc][cc]{$M$}
\psfrag{Th}[cc][cc]{$T_H$}
\psfrag{rho1}[cc][cc]{$\quad\rho(k_1)$}
\psfrag{rho2}[cc][cc]{$\quad\rho(k_2)$}
\scalebox{.53}
{$\begin{array}{cccc}
\raisebox{-0.44 \totalheight}
{\psfrag{r}[cc][cc]{$\quad\rho(k_1)$}
\psfrag{pf}[cc][cc]{$\slashchar{p}_2$}
\epsfig{file=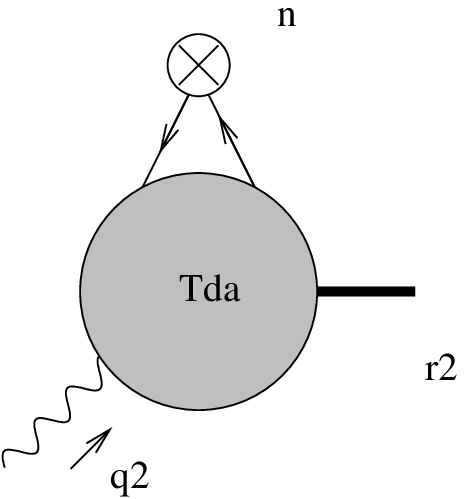,width=\scalingB}}&=&
\raisebox{-0.43\totalheight}{
\psfrag{Tda}[cc][cc]{$TDA_H$}
\epsfig{file=TDA.eps,width=\scalingC}}& \begin{array}{c}
\raisebox{0.51 \totalheight}
{\psfrag{r}[cc][cc]{$\quad\rho(k_1)$}
\psfrag{pf}[cc][cc]{$\slashchar{p}_2$}
}\\
\raisebox{-0.0 \totalheight}
{\psfrag{r}[cc][cc]{$\quad\rho(k_2)$}
\psfrag{pf}[cc][cc]{$\slashchar{p}_1$}
\epsfig{file=DA.eps,width=\scalingA}}
\end{array}
\end{array}$
}
\\
\hspace{-.3cm}
\vspace{.3cm}
\raisebox{-0.46 \totalheight}{\rm with}
\\
\vspace{.3cm}\hspace{-.3cm}
\psfrag{p}[cc][cc]{$\slashchar{p}_2$}
\psfrag{p2}[cc][cc]{$\slashchar{p}_2$}
\psfrag{q1}[cc][cc]{$q_1$}
\psfrag{q2}[cc][cc]{$q_2$}
\psfrag{Tda}[cc][cc]{$TDA_H$}
\psfrag{n}[cc][cc]{$\slashchar{p}_1$}
\scalebox{.5}
{
$\begin{array}{ccccccc}
\raisebox{-0.46 \totalheight}{\epsfig{file=TDA.eps,width=\scalingb}}
&=&
\raisebox{-0.46 \totalheight}{\epsfig{file=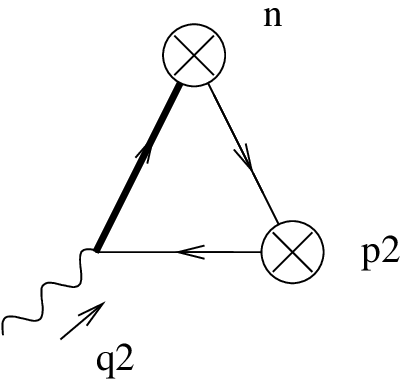,width=\scalingb}}
&+&
\raisebox{-0.46\totalheight}{\epsfig{file=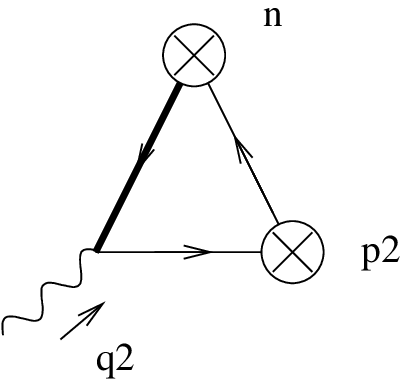,width=\scalingb}}
&+&
\raisebox{-0.42\totalheight}{\epsfig{file=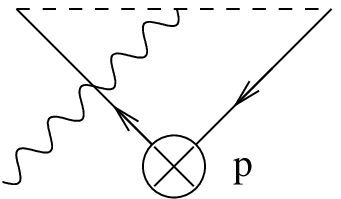,width=\scalingml}}
\end{array}
$}
\end{tabular}
\end{tabular}
\caption{a: Factorization of the amplitude in terms of a TDA. b: Perturbative TDA factorization.}
\label{FigTDAfact}
\end{wrapfigure}

\psfrag{p}[cc][cc]{$\slashchar{P}$}
\psfrag{q1}[cc][cc]{$q_1$}
\psfrag{q2}[cc][cc]{$\,1+\xi$}
\psfrag{1mxi}[cc][cc]{$\!1-\xi$}
\psfrag{xmxi}[cc][cc]{$x-\xi$}
\psfrag{xpxi}[cc][cc]{$\!\!x+\xi$}
\psfrag{CF}[cc][cc]{}
\psfrag{TDA}[cc][cc]{TDA}
\psfrag{r1}[cc][cc]{$\,\,\rho(k_1)$}
\psfrag{r2}[cc][cc]{$\,\, \rho(k_2)$}
\begin{wrapfigure}{r}{0.27\columnwidth}
\vspace{-1.6cm}
\begin{center}
\hspace{.6cm}
\scalebox{.65}{\raisebox{-0.46 \totalheight}{\epsfig{file=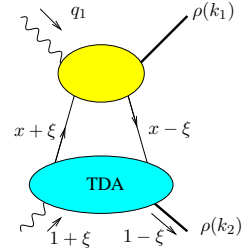,width=4.2cm}}}
\end{center}\vspace{-.3cm}
\caption{TDA kinematics.}
\label{FigTDAkin}
\vspace{-2cm}
\end{wrapfigure}
\vspace{-.0cm}
The amplitude 
$M= T^{\alpha\, \beta}p_{2\,\alpha} \,p_{1\,\beta}$ (\ref{resML}) can be interpreted in this limit
as
a convolution $M =   TDA \otimes CF \otimes DA,$ according to 
\begin{eqnarray*}
&&\hspace{-.6cm}T^{\alpha\,\beta}p_{2\,\alpha}p_{1\,\beta} =
-i \frac{C}{2}\int\limits_{-1}^1 dx\,\int\limits_0^1 dz_1\,
\left[\frac{1}{\bar z_1(x-\xi)} + \frac{1}{z_1(x+\xi)} \right]\, \phi(z_1) \\
&&\hspace{-.75cm} \times N_c \left[\Theta(1\ge x \ge \xi)\, \phi\left(\frac{x-\xi}{1-\xi} \right) -
\Theta(-\xi \ge x \ge -1) \, \phi\left( \frac{1+x}{1-\xi} \right) \right]\!,
\end{eqnarray*}
the TDA being defined through the usual GPD kinematics (see Fig.\ref{FigTDAkin}),
with $n_1 =(1+\xi) p_1$ and $n_2 =\frac{p_2}{1+\xi},$ defining
 $x, \xi$ as momentum fraction along $n_2.$
This factorisation (see Fig.\ref{FigTDAfact}a)  is proven at Born order by computing perturbatively the TDA $\gamma^* \to \rho_L^0$ defined as
\beas
&&\int\,\frac{dz^-}{2\pi s}\,e^{ix(n_2.z)}\,\langle \rho^q_L(k_2)|
\bar q(-z/2)\,\slashchar{n_1}\,\exp\{-ieQ_q\!\int\limits_{z/2}^{-z/2}\,
dy_\mu\,A^\mu(y) \}q(z/2)|\gamma^*(q_2)\rangle
\nonumber \\
&&= \frac{e\,Q_q\,f_\rho}{n_2^+}\,\frac{1}{Q_2^2}\,
\epsilon_\nu(q_2)\left((1+\xi)n_2^\nu +\frac{Q_2^2}{s(1+\xi)}n_1^\nu\right)\,T(x,\xi,t_{min})\;,
\eeas
where the QED Wilson line is explicitly indicated (QCD Wilson line
gives no contribution).
Since $Q_2^2$ hard, the TDA can be factorized (see Fig.\ref{FigTDAfact}b)  as 
\[
T(x,\xi,t_{min})\equiv N_c \left[\Theta(1\ge x \ge \xi)\, \phi\left(\frac{x-\xi}{1-\xi} \right) -
\Theta(-\xi \ge x \ge -1) \, \phi\left( \frac{1+x}{1-\xi} \right) \right]\,.
\]

\vspace{-.0cm}
\begin{wrapfigure}{r}{0.4\columnwidth}
\psfrag{p}[cc][cc]{$\,\, \slashchar{p}_2$}
\psfrag{p1}[cc][cc]{$\,\, \slashchar{p}_1$}
\psfrag{p2}[cc][cc]{$\,\, \slashchar{p}_2$}
\psfrag{q1}[cc][cc]{$q_1$}
\psfrag{Th}[cc][cc]{$T_H$}
\psfrag{n}[cc][cc]{$\slashchar{n}$}
\vspace{-.1cm}
\hspace{-.5cm}
\centerline{\scalebox{.42}
{
$
\begin{array}{ccccc}
\raisebox{-0.46 \totalheight}{\epsfig{file=THL.eps,width=\scalingl}}
&\, =&
\raisebox{-0.46 \totalheight}{\epsfig{file=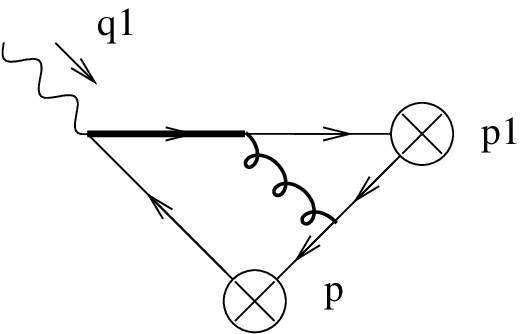,width=\scalingl}}
&\,+&
\raisebox{-0.46 \totalheight}{\epsfig{file=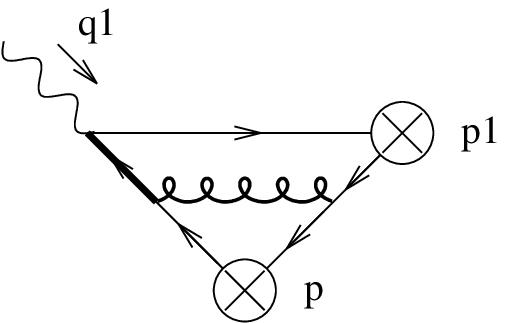,width=\scalingl}}
\\
&\,+&
\raisebox{-0.46\totalheight}{\epsfig{file=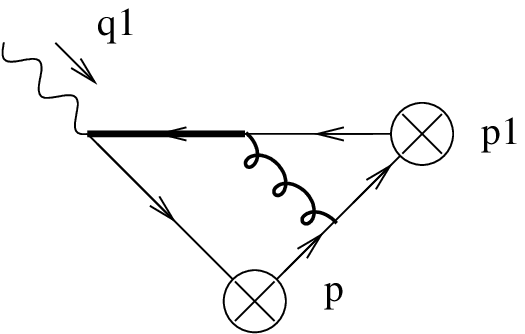,width=\scalingl}}
&\,+&
\raisebox{-0.46\totalheight}{\epsfig{file=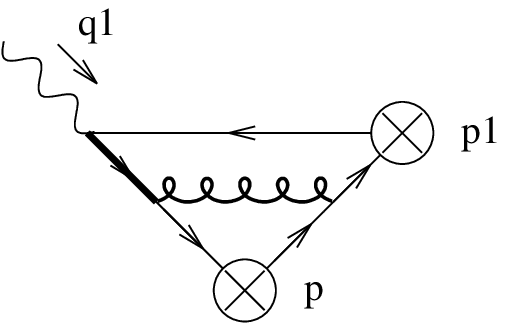,width=\scalingl}}
\end{array}
$}}
\caption{Hard part $T_H$ at lowest order.}\vspace{-1.6cm}
\label{HardTDA}
\end{wrapfigure}
\noi The Hard term reads, at Born order (see Fig.\ref{HardTDA}),
\begin{eqnarray*}
&&\hspace{-.65cm}T_H(z_1,x)=-i\,f_\rho\,g^2\,e\,Q_q\,\frac{C_F\,\phi(z_1)}{2\,N_c\,Q_1^2}
\epsilon^\mu(q_1)\nonumber \\
&&\hspace{-.71cm} \times\!\left(\!2\xi\,n_{2\,\mu} \! +\!\frac{1}{1+\xi}n_{1\,\mu}
\! \right)  \!\!
 \left[\frac{1}{z_1(x+\xi-i\epsilon)} \!+\! \frac{1}{\bar
z_1(x-\xi+i\epsilon)}    \right]\,,
\end{eqnarray*}

\section{Computation at large $W^2$}

\noi The dynamics of QCD
in the perturbative  Regge limit\cite{slidesrev} is governed by gluons. 
BFKL 
enhancement effects are expected to be important at large  rapidity. 
The exclusive process (\ref{processgg}) tests this limit\cite{PSW,EPSW,SSW},
 for both $Q_i^2$   {\it hard} 
and of the {\it same order} (to suppress collinear dynamics à la DGLAP\cite{dglap} and ERBL\cite{ERBL}),
giving access to the full non-forward  Pomeron structure, in relation with 
saturation studies, where a full impact parameter picture
is needed. 
 Increasing  $s_{\gamma^*\gamma^*}$ for fixed values $Q_1^2$
and $Q_2^2$ causes transition from the linear to non-linear (saturated) regime. 


\vspace{-.35cm}
\begin{figure}[h]
\scalebox{.95}{
\begin{picture}(600,260)
\put(-105,129){\epsfxsize=\wids{\centerline{\epsfbox{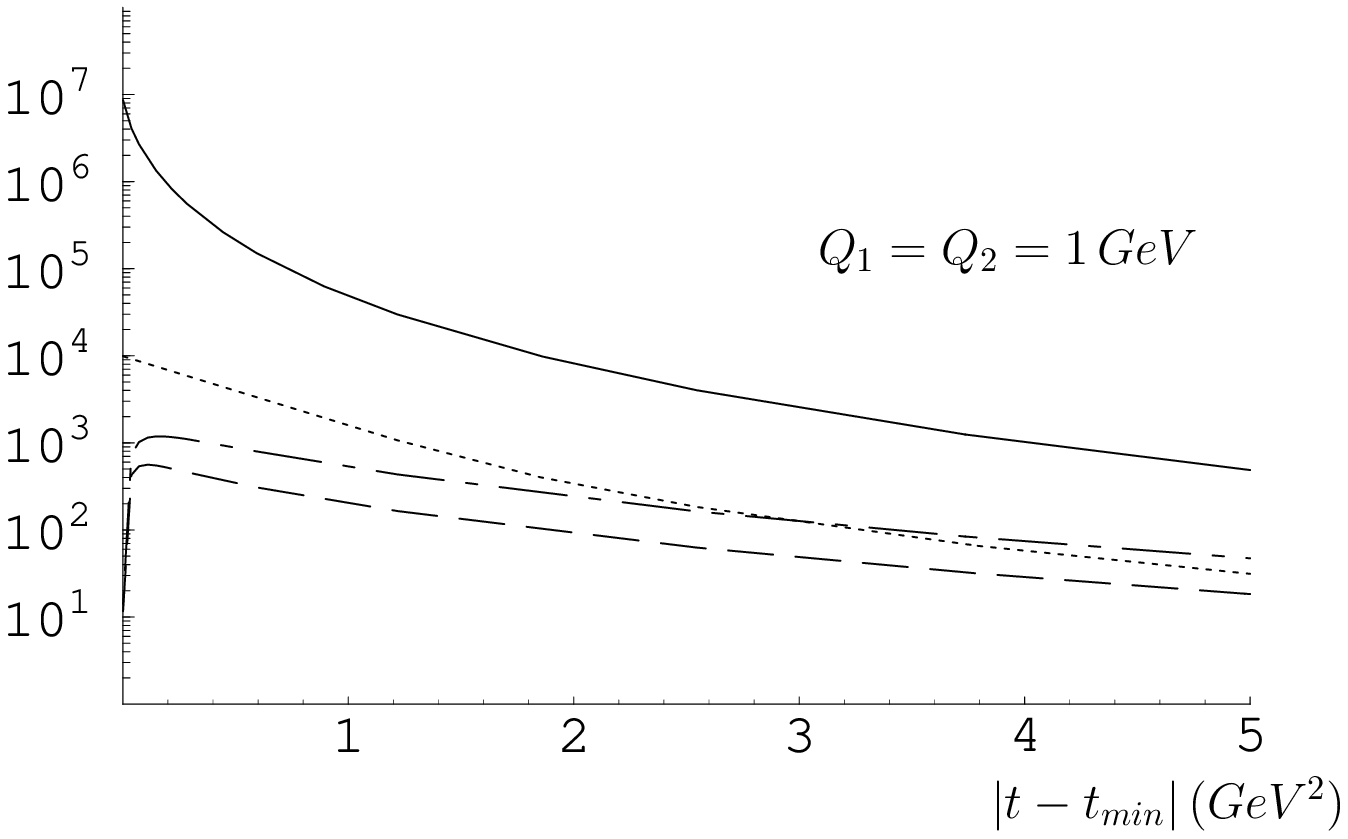}}}}
\put(105,130){\epsfxsize=\wids{\centerline{\epsfbox{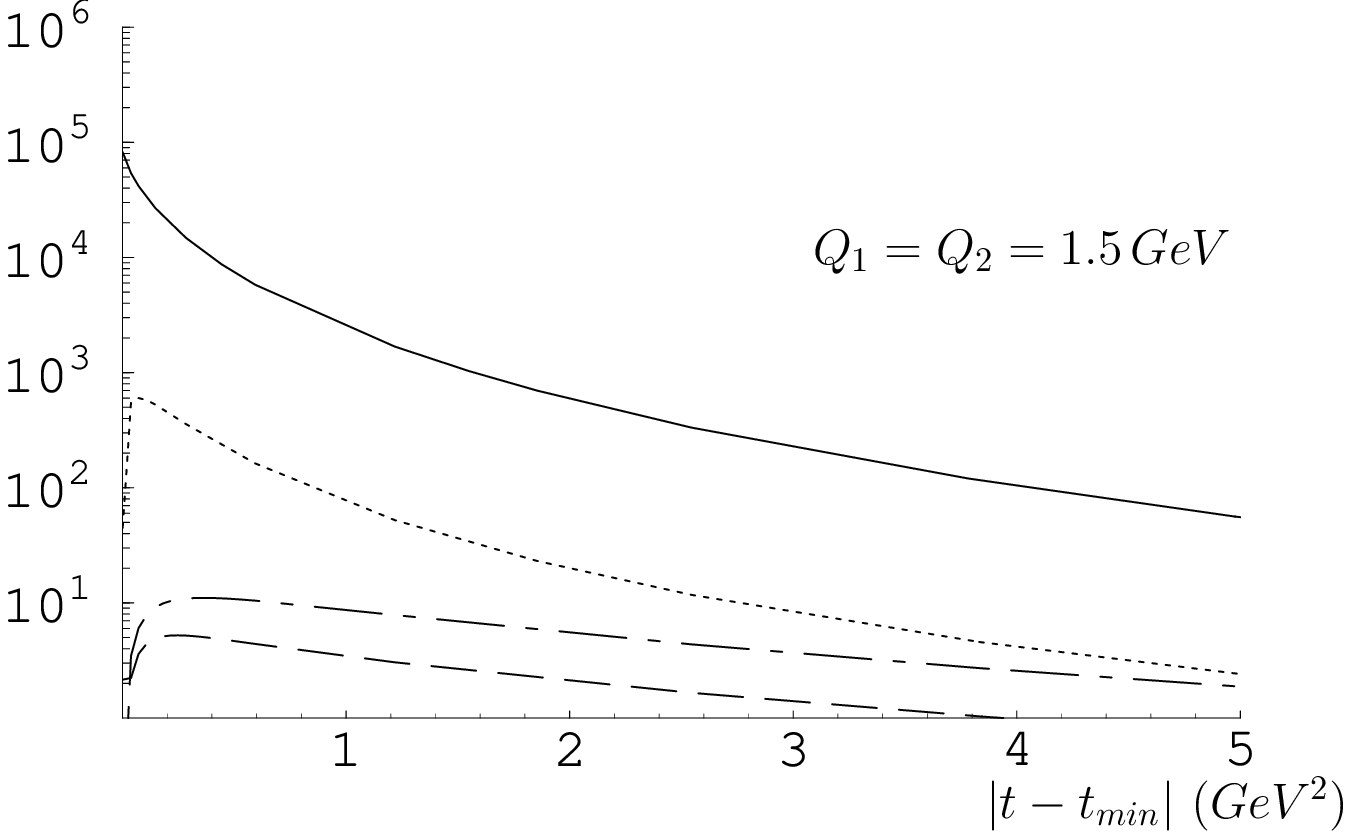}}}}\put(-105,2.5){\epsfxsize=\wids{\centerline{\epsfbox{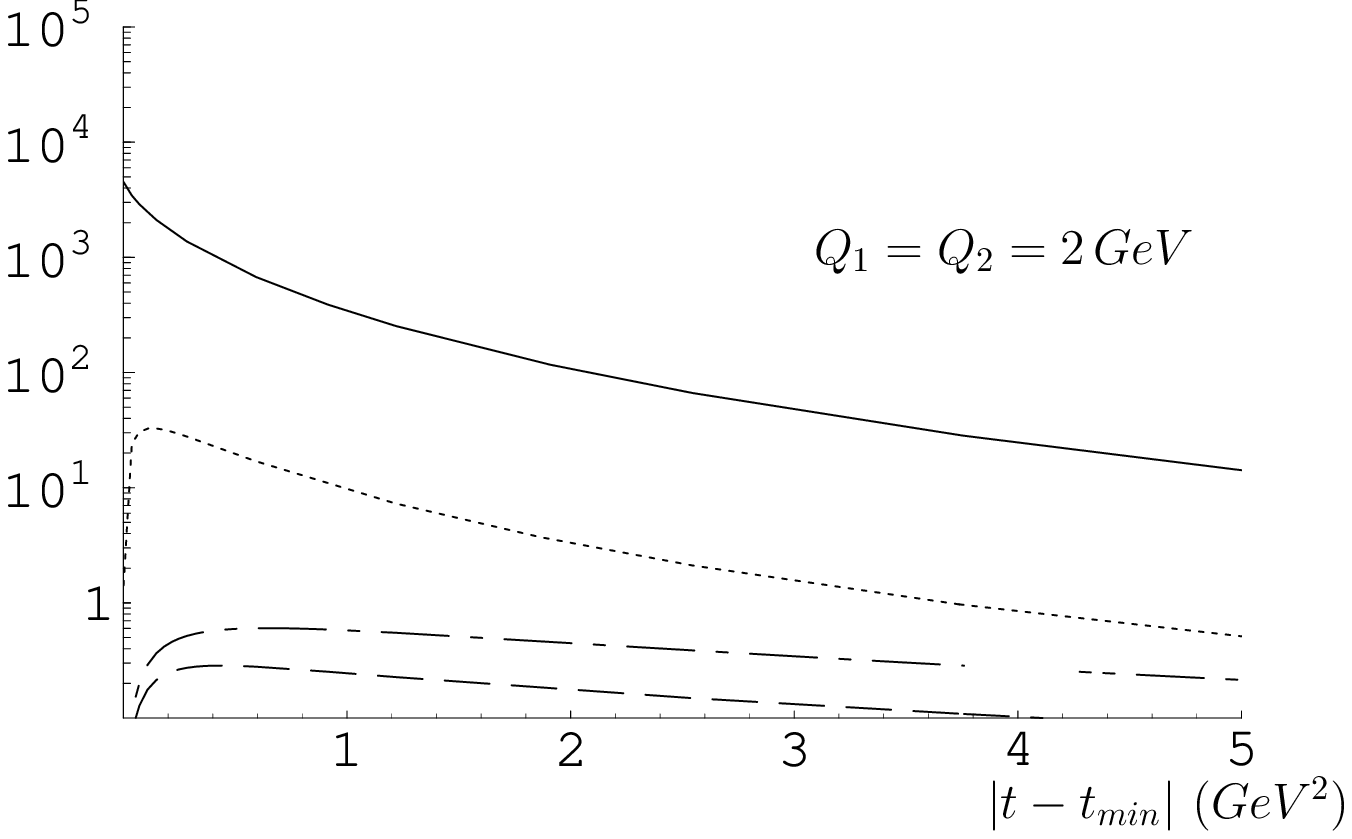}}}}
\put(36,223){\tiny$LL$}
\put(36,196){\tiny$LT$}
\put(36,185){\tiny$T\neq T'$}
\put(36,170){\tiny$T= T'$}

\put(246,219){\tiny$LL$}
\put(246,192){\tiny$LT$}
\put(246,168){\tiny$T\neq T'$}
\put(246,149){\tiny$T= T'$}

\put(36,90){\tiny$LL$}
\put(36,62){\tiny$LT$}
\put(36,38){\tiny$T\neq T'$}
\put(36,19){\tiny$T= T'$}

\put(100,5){\epsfxsize=\wids{\centerline{\epsfbox{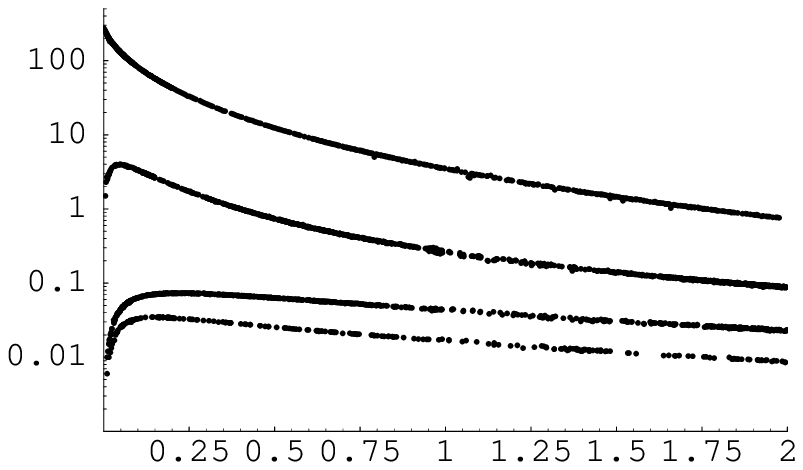}}}}
\put(250,110){\tiny $LL$}
\put(250,82){\tiny$LT$}
\put(250,58){\tiny$T\neq T'$}
\put(250,31){\tiny$T= T'$}
\put(350,3){\scalebox{.6}{ $|t-t_{min}|  \, (GeV^2)$}}
\put(225,125){\scalebox{.65}{$\frac{d\sigma^{e^+ e^- \to \, e^+ e^- \rho_L \rho_L}}{dt} (fb/GeV^2)$}}

\put(120,230){(a)}
\put(330,230){(b)}
\put(120,110){(c)}
\put(330,110){(d)}
\end{picture}
}\vspace{-.45cm}
\caption{$\gamma^*_{L,T}\gamma^*_{L,T} \to  \rho_L^0  \;\rho_L^0$  (a,b,c) and $e^+e^- \to e^+e^- \rho_L^0  \;\rho_L^0$ (d) differential cross-sections.}\vspace{-.3cm}
\label{FigDiff}
\end{figure}

\noi When $s_{\gamma^*\gamma^*}
  \gg -t, Q_1^2,Q_2^2,$ we rely on the impact representation
which reads, at Born order,\vspace{-.1cm}
$$
{\cal M} = is\;\int\;\frac{d^2\,\kb}{(2\pi)^4\kb^2\,(\rb -\kb)^2}
{\cal J}^{\gamma^*_{L,T}(q_1) \to \rho^0_L(k_1)}(\kb,\rb -\kb)\;
{\cal J}^{\gamma^*_{L,T}(q_2) \to \rho^0_L(k_2)}(-\kb,-\rb +\kb)
$$
where the impact factors ${\cal J}^{\gamma^*_{L,T}}$ are rational functions of  the transverse momenta $(\kb,\rb)$.
The  2-d 

\noi integration is treated analytically, relying 
 on conformal transformations in the transverse momentum plane. 
The integrations over momentum fractions $z_1$ and $z_2$ (hidden in ${\cal J}$) are performed numerically.
We use $Q_1 Q_2$ as a scale for $\alpha_S$.
As displayed in Fig.\ref{FigDiff}a,b,c,
cross-sections are strongly peaked  at small $Q^2$ and small $t,$
and longitudinally polarized photons dominates.
The \nopagebreak[4] non-forward Born order cross-section for
 $e^+e^- \to e^+e^- \rho_L^0  \;\rho_L^0$ is obtained with the help of the 
 equivalent photon approximation. Defining 
 $y_{i}$ as the longitudinal momentum fractions of the bremsstrahlung photons, 
one finds that $\sigma^{e^+ e^- \to e^+ e^- \rho_L \rho_L}$  gets its main contribution from  the low $y$ and $Q^2$ region,  which is the very forward region.
At 
ILC, 
${\sqrt s_{e^+e^-}}   = 500\, {\rm GeV},$ 
 with $125 \,{\rm fb^{-1}}$ per year. The measurement seems feasible since
each detector design includes a very forward electromagnetic calorimeter for luminosity measurement,
with tagging angle for outgoing leptons down to 5 mrad.
In Fig.\ref{FigDiff}d, we display our results within
the Large Detector Concept.
We obtain
$\sigma^{tot}= 34.1 \,{\rm fb}$ 
and $4.3 \,10^3$ events per year.
The  LL BFKL enhancement is enormous but not trustable, since it is well known that NLL  BFKL
is far below LL. Work  to implement  {\it resummed} LL  BFKL effects\cite{Khoze} 
is in progress, with results in accordance with the NLL based one\cite{IvanovPapa}. The obtained enhancement  is less dramatic ($\sim 5$) than with LL BFKL, but {\it still visible}.

\vspace{-.4cm}
\section*{Aknowledgements}

L.Sz. is supported by the Polish Grant 1 P03B 028 28. He
 is a Visiting Fellow of FNRS \nolinebreak[4] (Belgium).\linebreak

\vspace{-.9cm}
\begin{footnotesize}
\bibliographystyle{blois07}

\begin{thebibliography}{99}



\bibitem{bfkl} E.A.~Kuraev,  L.N.~Lipatov and V.S.~Fadin,
Phys.\ Lett.\ B {\bf 60}, 50-52 (1975);
Sov.\ Phys.\ JETP {\bf 44}, 443-451 (1976) ;
 Sov.\ Phys.\ JETP {\bf 45}, 199-204 (1977) ;
Ya.Ya.~Balitskii and L.N.~Lipatov,
Sov.\ J.\ Nucl.\ Phys. {\bf 28}, 822-829 (1978).


 \bibitem{DM}
D.~M{\"u}ller, D.~Robaschik, B.~Geyer, F.~M.~Dittes, J.~Ho\v{r}ej\v{s}i,
Fortsch.\ Phys.\  {\bf 42},  101 (1994);

\bibitem{GPD}
X.~Ji,
Phys.\ Rev.\ Lett.\  {\bf 78}, 610 (1997);  Phys.\
  Rev.\ D {\bf 55}, 7114 (1997);
A.~V.~Radyushkin,
Phys.\ Rev.\ D {\bf 56}, 5524 (1997);
For a review, see M.~Diehl,
Phys.\ Rept.\  {\bf 388},  41 (2003) and
 A.~V.~Belitsky and A.~V.~Radyushkin,
Phys.\ Rept.\  {\bf 418}, 1 (2005).



 \bibitem{GDA}
M.~Diehl, T.~Gousset, B.~Pire and O.~Teryaev,
Phys.\ Rev.\ Lett.\  {\bf 81}, 1782  (1998);
M.~Diehl, T.~Gousset and B.~Pire,
Phys.\ Rev.\ D {\bf 62}, 073014 (2000);
I.~V.~Anikin, B.~Pire and O.~V.~Teryaev,
Phys.\ Rev.\ D {\bf 69}, 014018  (2004).

\bibitem{TDA}
  B.~Pire and L.~Szymanowski,
  Phys.\ Rev.\ D {\bf 71}, 111501 (2005);
  Phys.\ Lett.\ B {\bf 622}, 83 (2005);
  J.~P.~Lansberg, B.~Pire and L.~Szymanowski,
  Phys.\ Rev.\ D {\bf 73},  074014 (2006);
 Phys.\ Rev.\ D {\bf 75},  074004 (2006) and arXiv:0709.2567 [hep-ph].


\bibitem{gdatda} B.~Pire, M.~Segond,
 L.~Szymanowski and  S.~Wallon,
Phys.\ Lett.\ B {\bf 39}, 642-651 (2006);
  PoS D {\bf IFF2006}, 034 (2006).

\bibitem{BLphysrev24}
  S.~J.~Brodsky and G.~P.~Lepage,
  Phys.\ Rev.\ D {\bf 24}, 1808 (1981).

\bibitem{DFKV}
  M.~Diehl, T.~Feldmann, P.~Kroll and C.~Vogt,
  Phys.\ Rev.\ D {\bf 61}, 074029 (2000).

\bibitem{slidesrev} See talk "Perturbative QCD in the Regge limit: Prospects at ILC",  arXiv:0710.0833 [hep-ph],
slides: https://indico.desy.de/conferenceDisplay.py?confId=372 

\bibitem{SSW}
 M.~Segond, L.~Szymanowski and S.~Wallon,
 Eur.\ Phys.\ J.\  C {\bf 52}, 93 (2007)
  [arXiv:hep-ph/0703166].

\bibitem{PSW}
  B.~Pire, L.~Szymanowski and S.~Wallon,
  Eur.\ Phys.\ J.\ C {\bf 44},  545 (2005);
LCWS 04, Paris, France, 19-24 Apr 2004,
Published in "Paris 2004, Linear colliders, vol. 1," 335-340;
Nucl.\ Phys.\ A {\bf 755} (2005) 626-629;
EDS05, Blois, France, 15-20 May 2005, "Towards High Energy Frontiers",
eds. M.~Haguenauer et al., 373-376.

\bibitem{EPSW}
  R.~Enberg, B.~Pire, L.~Szymanowski and S.~Wallon,
  Eur.\ Phys.\ J.\ C {\bf 45},  759 (2006) [Erratum-ibid.\  C {\bf 51}, 1015 (2007)]; 
Acta Phys.\ Polon.\  B {\bf 37}, 847 (2006).



\bibitem{dglap}
V.~N.~Gribov and L.~N.~Lipatov,
Yad.\ Fiz.\  {\bf 15},  (1972) 781
[Sov.\ J.\ Nucl.\ Phys.\  {\bf 15}, 438 (1972)];
G.~Altarelli and G.~Parisi,
Nucl.\ Phys.\ B {\bf 126},  298 (1977);
Y.~L.~Dokshitzer,
Sov.\ Phys.\ JETP {\bf 46},  641 (1977)
[Zh.\ Eksp.\ Teor.\ Fiz.\  {\bf 73},  1216 (1977)].




\bibitem{ERBL}
G.P. Lepage and S.J. Brodsky, Phys.\ Lett.\ B {\bf 87},
359 (1979); A.V. Efremov and A.V. Radyushkin, Phys.\ Lett.\ B {\bf 94}, 245
 (1980).


\bibitem{Khoze}
V.~A.~Khoze, A.~D.~Martin, M.~G.~Ryskin and W.~J.~Stirling,
Phys.\ Rev.\ D {\bf 70},   074013 (2004).


\bibitem{IvanovPapa} D.~Y.~Ivanov and A.~Papa,
  Nucl.\ Phys.\  B {\bf 732},  183 (2006)
and Eur. \ Phys. \ J. \ C {\bf 49} 947 (2007).


\nopagebreak[4]
\bibitem{slidestalk1} slides: https://indico.desy.de/conferenceDisplay.py?confId=372\#9





\end{thebibliography}

\end{footnotesize}

\end{document}